\documentstyle[11pt,mypasp,twoside,epsf]{article}
\begin{document}
\title{Objective-Prism Emission-Line Searches for Active Galactic Nuclei}
\author{Caryl Gronwall}
\affil{Dept. of Physics \& Astronomy, Johns Hopkins University, Baltimore MD 21218}
\author{Vicki L. Sarajedini}
\affil{Dept. of Astronomy, University of Florida, Gainesville, FL 32611}
\author{John J. Salzer}
\affil{Astronomy Dept., Wesleyan University, Middletown, CT 06459}

\begin{abstract}
Objective-prism surveys for UV-excess and emission-line objects,
especially the First and Second Byurakan Surveys, have been central to 
the study of active galactic nuclei (AGNs).  We review 
previous line-selected surveys for AGNs and discuss their contribution
to our understanding of the AGN phenomena.  In addition, we present 
results from the KPNO International Spectroscopic Survey, a modern digital
objective-prism survey for emission line objects.  This survey is
discovering substantial numbers of new AGNs, in particular low-luminosity
AGNs and LINERs.
\end{abstract}

\section{Introduction}
Objective-prism surveys revolutionized our knowledge of active
galactic nuclei (AGNs) in the local universe.  In particular,
many of the known nearby Seyfert galaxies were discovered by
Markarian in the First Byurakan Survey (FBS; see Khachikian this volume.)
Galaxies were selected for inclusion in the FBS based on the
presence of a UV excess.  This selection criterion is biased against
the discovery of more heavily reddened AGN such as Seyfert 2's and 
LINER galaxies.  A number of line-selected surveys have been carried
out over the past 20 years in order to select well-defined samples
of local AGNs.  In this paper, we review existing line-selected surveys
for local AGNs carried out using objective-prism techniques (see Stepanian,
this volume, for a discussion of the UV+line selected
Second Byurakan Survey).  We then
present a discussion of a new, modern digital objective-prism survey for
emission-line galaxies (ELGs), 
the KPNO International Spectroscopic Survey (KISS).
KISS probes several magnitudes deeper than existing surveys, and due
to its selection via H$\alpha$ emission, is less biased against
Seyfert 2 and LINER galaxies than previous surveys.  KISS provides
one of the best, well-defined samples of local AGNs for comparison
to higher redshift studies.

\section{Previous Line-Selected Surveys}

The use of line-selected objective-prism surveys to detect
active and star-forming galaxies in the local universe has 
a long and fruitful history (e.g., Smith 1975; Smith et al.~1976). 
The University of Michigan (UM) survey
(MacAlpine et al. 1977; MacAlpine \& Lewis 1981)
selected objects via their [OIII]$\lambda5007$ emission line. 
349 ELGs were detected in 667 sq. degrees.  
This is a number density of 0.52 per square degree. In Lists
4 \& 5 of the UM survey (for which complete follow-up spectroscopy
is available from Salzer et al. 1989) 166 ELGs are found in 325 sq. degrees.
Approximately 10\% of the ELGs detected in the UM survey are AGNs:
9 Seyfert 1's and 7 Seyfert 2's.
Wasilewski (1983) also utilized the [OIII]-line as his
selection criterion and found 96 ELGs in 825 sq. degrees, or 0.18 per
square degree.  Wasilewski found that 8\% are
AGNs with 1 Seyfert 1 and 7 Seyfert 2 galaxies being detected.
The Case survey (Pesch \& Sanduleak 1983; Stephenson et al. 1992)
used two selection criteria, both UV-excess and [OIII]-line emission
and cataloged 1440 ELGs in 1551 sq. degrees or 0.94 per sq. degree.
Salzer et al. (1995) obtained follow-up spectroscopy for a complete
subsample of 176 of the Case galaxies and found
that only 6\% are AGNs: 2 Seyfert 1's, 7 Seyfert 2's, and 2 LINERs.  
Most recently the Universidad Complutense de Madrid (UCM; Zamaraono
et al. 1994, 1996) survey utilized selection via H$\alpha$-emission
to discover 263 ELGs in 471 sq. degrees or 0.56 per square degree.
Follow-up spectroscopy by Gallego et al. (1997) found that 14 
(or 5\%) of these are AGNs, including 5 Seyfert 1's and 9 Seyfert 2's.
The completeness limits of all of the above surveys range from
$B = 15$ to 17.

Most of the previous line-selected surveys discussed above
have been done in the blue (selecting via [OIII]$\lambda5007$ emission)
and all used photographic plates.  There are a number of biases introduced by
these methods.  First, selection in the blue limits the sensitivity
of the surveys to more heavily reddened Seyfert 2's and low-ionization
LINER galaxies.  Second, because the standard IIIa-J emulsions used
in most photographic surveys cut off at $\sim 5350$ \AA, the redshift
depth of these surveys is limited to $z \sim 0.065$.  A modern, digital
survey using CCDs to substantially improve survey depth (to fainter magnitudes
and higher redshifts) plus selecting via a redder line (H$\alpha$)
would greatly reduce these biases.  Such a survey is now available,
the KPNO International Spectroscopic Survey (KISS).

\section{The KPNO International Spectroscopic Survey}

The KPNO International Spectroscopic Survey (KISS) was initiated in 1994 as
a collaborative effort between astronomers from Russia, Ukraine, and the
US who shared a common interest in the study of galaxian activity.  The
survey was envisioned as being the ``next generation Markarian survey" 
for active and star-forming galaxies.  By combining the traditional 
objective-prism survey method for emission line detection with 
modern detector technology and computer-based analysis, KISS is able to
discover Seyfert Galaxies (Sy 1/Sy 2), LINER's,
Starburst Nucleus Galaxies (SBNs), 
HII Galaxies, and 
Blue Compact Dwarfs (BCDs) at least 2 magnitudes fainter than 
previous work.  The survey is thus a unique resource for probing
the nature of star-forming and active galaxies in the nearby universe.
A full description of the survey can be found in Salzer et al. (2000, 2001).

The KISS survey is conducted with CCD detectors on the 24-inch Burrell 
Schmidt on Kitt Peak.  For the first survey strip, a 2048$^2$ CCD was
used, which afforded a field-of-view of 70 arcmin (1.18$^\circ$) square, 
with a scale of 2.03 arcsec/pixel.  This one-degree-wide strip was chosen 
to coincide with the Century Redshift survey (Geller et 
al.~1997) covering a strip at a constant declination of 
29$^\circ$ from RA = 12$^h$ 15$^m$ to 17$^h$ which encompasses an 
area of 62 deg$^2$.  The second survey strip 
goes through the Bo\"{o}tes void at a constant declination of
43$^\circ$ from RA = 12$^h$ to 16$^h$ 15$^m$ covering an
area of 66 deg $^2$. For the second strip
a $2048 \times 4096$ CCD with 1.45 arcsec/pixel resolution was used.
The KISS survey data consists of broadband $B$ and $V$ images (for
astrometric and photometric calibration),  and
spectral (objective prism) data.
The spectra were taken in the red 
(using a blocking filter to restrict the wavelength range
to 6400-7200 \AA) to detect H$\alpha$ emission.
One of the important features of the KISS project is its objective criteria:
the search and measurement process is carried out entirely by computer.
A complete package of IRAF scripts and executables analyzes every source in
the field (typically 5000 -- 7000 objects) automatically; the output of
these routines is a catalog of photometric and astrometric data, and 
an extracted spectrum for every object in the field.  In addition,
a list of ELG candidates is produced.

Our survey technique has proven to be quite successful.
In our first red spectral strip we have cataloged 1128 ELG
candidates.  This is about {\it 18.1 per square 
degree}.  In the second survey strip, we have detected 1030 ELG
candidates or 15.6 per square degree.
For comparison, the entire Markarian survey cataloged 1500 
UV-excess galaxies (0.1 galaxy per square degree), and had a completeness 
limit of m$_B$ = 15.2 (Mazzarella \& Balzano 1986), while the deeper
UM (median m$_B$ = 16.9; MacAlpine et al.~1977) and UCM 
(median m$_B$=16.5; Zamarano et al.~1994) surveys both detected 
only 0.5 ELGs per square degree.  

\begin{figure}[hp]
\plotone{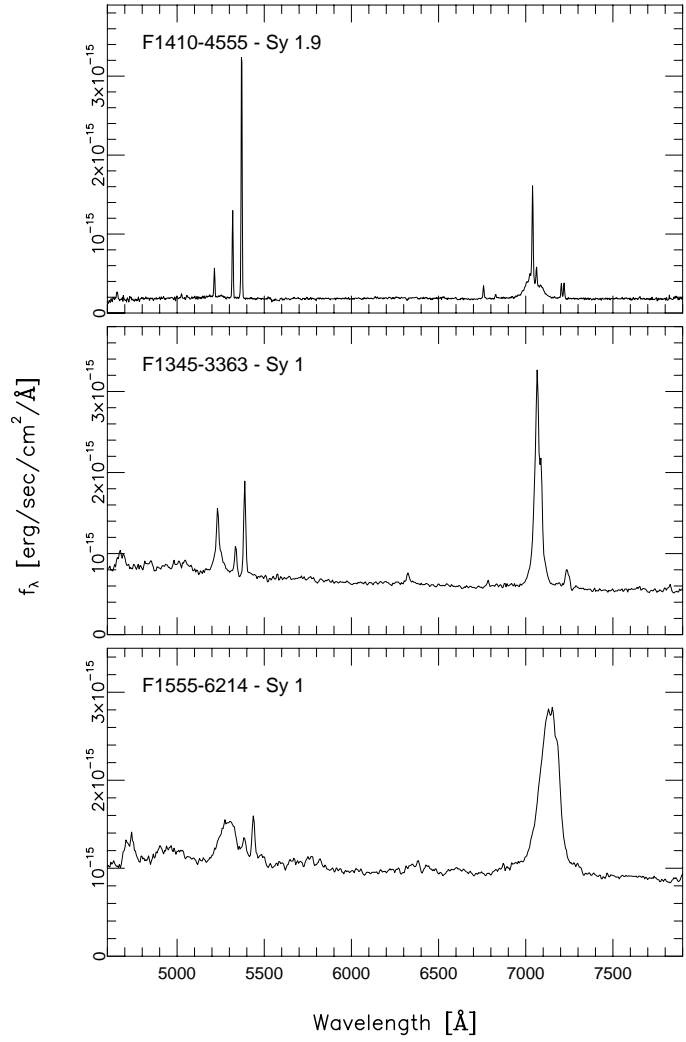}
\caption{Representative spectra of KISS Seyfert 1 galaxies.}
\end{figure}
\begin{figure}[hp]
\plotone{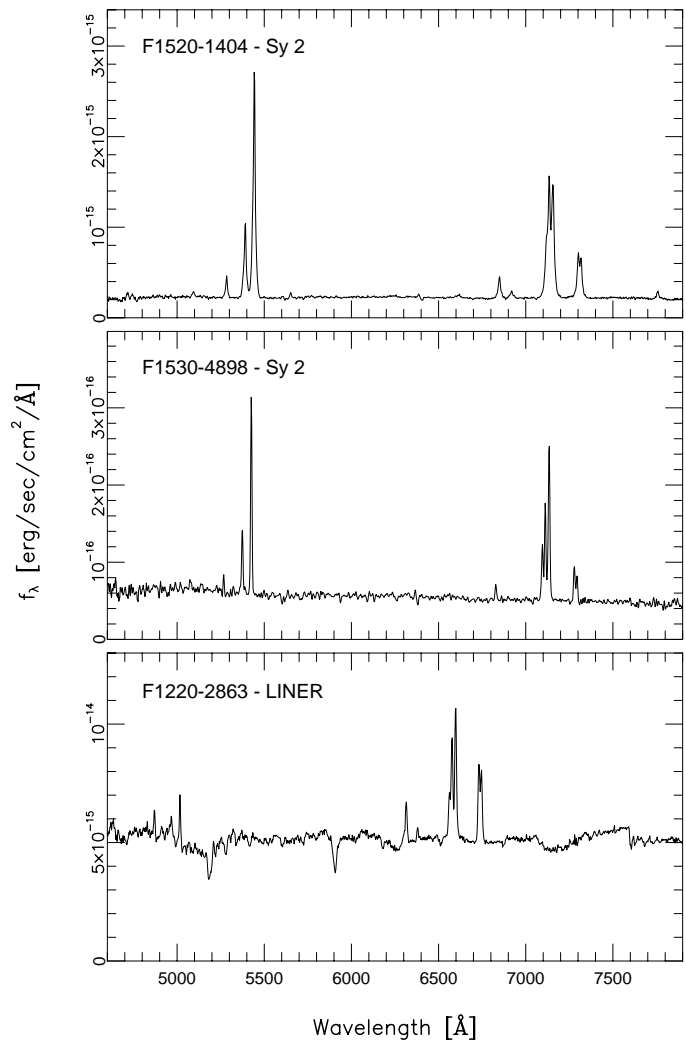}
\caption{Representative spectra of KISS Seyfert 2 and LINER galaxies.}
\end{figure}

\begin{figure}[t]
\plotone{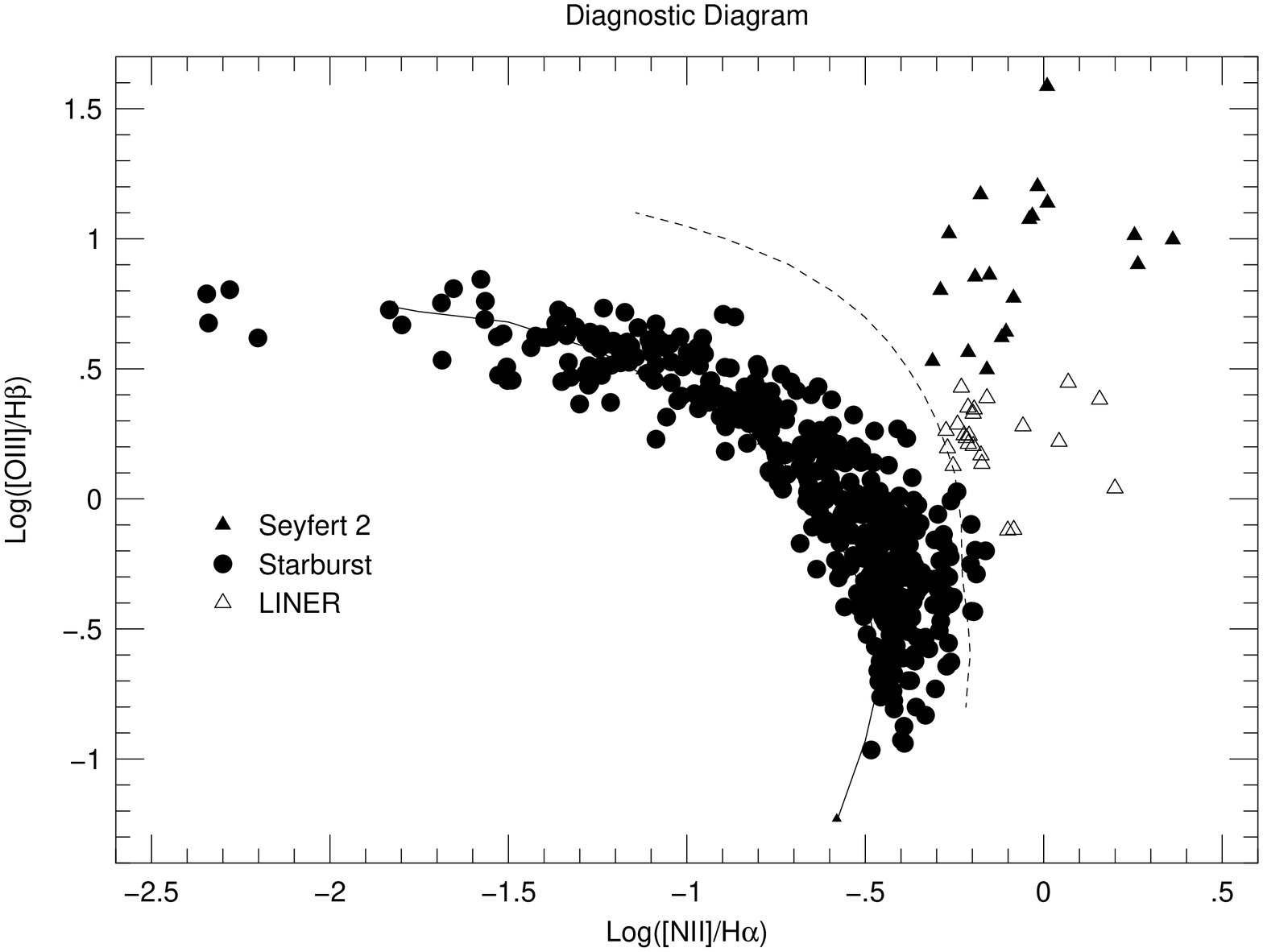}
\caption{Line diagnostic diagram plotting the logarithm
of [O~III]$\lambda$5007/H$\beta$ against the logarithm of 
[N~II]$\lambda$6583/H$\alpha$.  Solid circles represent star-forming
ELGs, open triangles are 
LINERS, and the solid triangles are Seyfert 2 galaxies.  The solid
line represents an HII model sequence at various metallicities from 0.1
Z$_{\odot}$ at the upper left to 2 Z$_{\odot}$ at the lower right 
(Dopita \& Evans 1986).}
\end{figure}

\subsection{Spectroscopic Follow-up}

We have obtained follow-up spectroscopy for 725 of 1128 ELG candidates
selected via H$\alpha$\ emission from the first survey strip
using the WIYN 3.5-m, the
KPNO 2.1-m, the
MDM 2.4-m, the APO 3.5-m, and the Lick 3-m telescopes as
well as the Hobby Eberly Telescope.
These spectra provide redshifts, H$\alpha$\
equivalent widths, and line fluxes for various important emission
lines including H$\beta$, [O~III]$\lambda\lambda$4959,5007,
                HeI$\lambda$5876, [O~I]$\lambda$6300, 
                H$\alpha$, [N~II]$\lambda\lambda$6548,6583, and 
                [S~II]$\lambda\lambda$6717,6731.
Measurements of the Balmer decrement allow us to directly measure the
extinction ($A_V$) for each of these galaxies.  The follow-up
spectra also allow us to identify AGNs in our sample:  Seyfert 1s
are identified via their broad permitted emission lines, while
Seyfert 2's and LINERs are distinguished primarily 
via the [OIII]/H$\beta$ 
line ratio with LINERs having [OIII]/H$\beta$ $<$ 3 (e.g., Veilleux \&
Osterbrock 1987).
Representative spectra of AGN in the KISS sample are shown
in Figures 1 \& 2.

We find that 91\% (663) of the KISS candidate ELGs are confirmed
emission-lines galaxies.  
About 11\% of the galaxies
are active galactic nuclei with 67 (10 Seyfert 1's,
19 Seyfert 2's, and 38 LINER galaxies) detected via H$\alpha$
with redshifts less than 0.095, and an additional 10 (1 Seyfert 1,
5 Seyfert 2's and 4 QSO's) detected at higher redshifts where
another line has redshifted into our filter.
A line diagnostic diagram plotting the logarithm
of [O~III]$\lambda$5007/H$\beta$ against the logarithm of 
[N~II]$\lambda$6583/H$\alpha$ is shown in Figure 3.  The solid line 
represents the HII sequence with low metallicity, high-ionization ELGs
in the upper left and high metallicity, low-ionization objects in the
lower right part of the sequence.  The star-forming ELGs follow the
HII sequence allowing us to classify them from BCDs in the
upper left to starburst nuclei in the lower right.  AGNs are clearly
separated from star-forming galaxies.  We find 
many more high metallicity, low-ionization objects than found with
traditional [O~III]$\lambda$5007-selected surveys.  The data
make it clear that  H$\alpha$-selected surveys are much more effective
at detecting
the full range of ionization and metallicity present in
star-forming galaxies than [O~III] or [O~II] observations.

\subsection{Properties of KISS AGN}

\begin{figure}[t]
\plotone{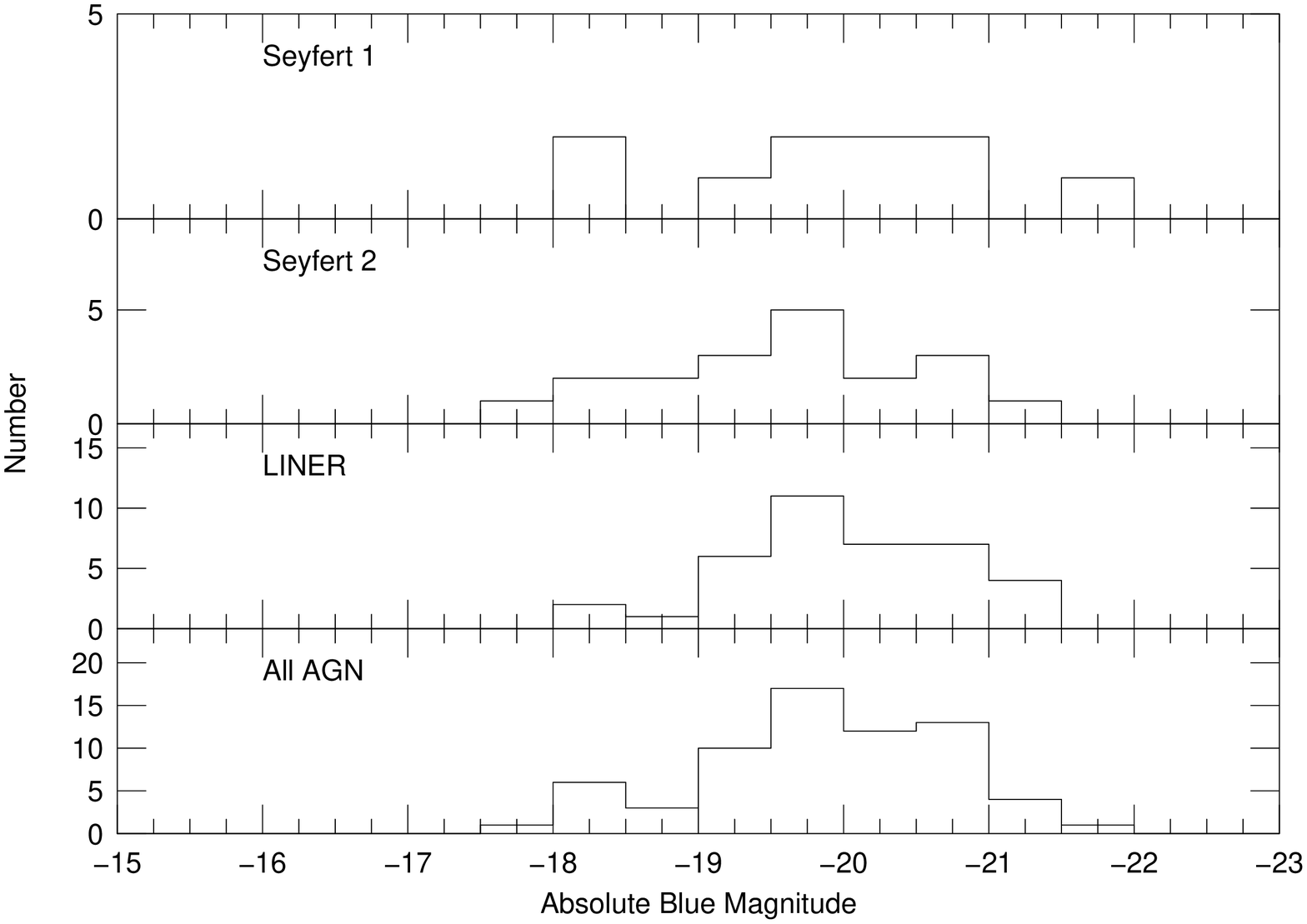}
\caption{$B - V$ colors of Seyfert 1's, Seyfert 2's, LINER's and all KISS AGN.
}
\end{figure}

\begin{figure}[t]
\plotone{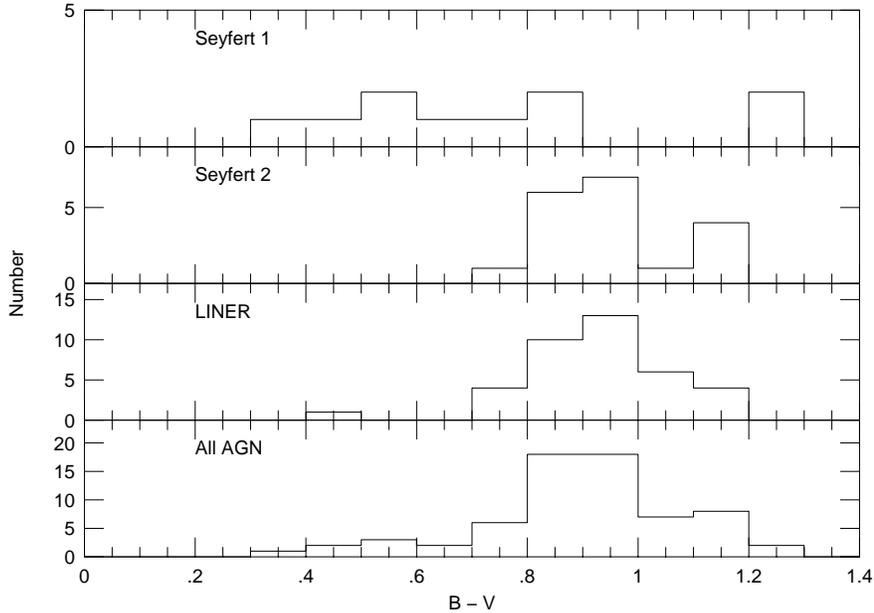}
\caption{$B - V$ colors of Seyfert 1's, Seyfert 2's, LINER's and all KISS AGN.}
\end{figure}

We will concentrate here on the 67 H$\alpha$-selected AGN which
constitute a well-defined local sample of AGN.  Note that because
our survey has a sharp upper-wavelength cutoff due to the blocking filter
used, the sample is {\it volume-limited} for the more luminous objects
and all but 8 of the AGNs fall in this category.  Figure 4 shows the
integrated absolute magnitude distributions of the 
AGNs in the local KISS sample.  KISS detects AGNs with $M_B$ ranging
from -21.8 to -18 (assuming $H_0 = 75$ km s$^{-1}$ Mpc$^{-1}$).  
The Seyfert 1, Seyfert 2, and LINER
luminosity distributions look similar.  Figure 5 shows the
$B - V$ color distributions of the KISS AGNs.  As expected, 
Seyfert 2 and LINER
distributions are significantly redder (median $B - V$ = 0.92 and 0.93
respectively) than the Seyfert 1 color distribution (median $B - V = 0.70$).
Because of their redder colors, an H$\alpha$-selected
survey is much more sensitive to Seyfert 2's and LINERs than 
a blue (e.g., UV excess or [OIII]-selected) survey.  This is borne out by
comparing the 2:1 LINER/Sy 2 ratio in the KISS survey to the 
complete lack of LINERs discovered in the [OIII]-selected surveys discussed
in Section 2.

We have also calculated the $B$-band luminosity function for
the Seyfert galaxies in our sample.  Because our direct images
lack the angular resolution required to resolve the nucleus, this
is an {\it integrated\/} luminosity function.  However, it is
directly comparable to the integrated local Seyfert luminosity
function of Huchra \& Burg (1992).  Figure 6 shows the blue luminosity
function for both 
the KISS sample and that of Huchra \& Burg.  The two
LFs are consistent, with the KISS LF extending to
fainter luminosities.  Note that for this calculation
we have not accounted for the possibility that there are AGNs in
the fraction of our sample for which we have no follow-up spectra.
Because we have preferentially obtained spectra for the more
luminous galaxies in our sample, we expect that such a correction
would be small, but it could increase our volume densities slightly.
Further discussion of the LF and its implications for AGN evolution
when compared to other local and higher redshift AGN LFs will
be discussed in a future paper (Sarajedini, Gronwall \& Salzer 2002).

\begin{figure}[t]
\plotone{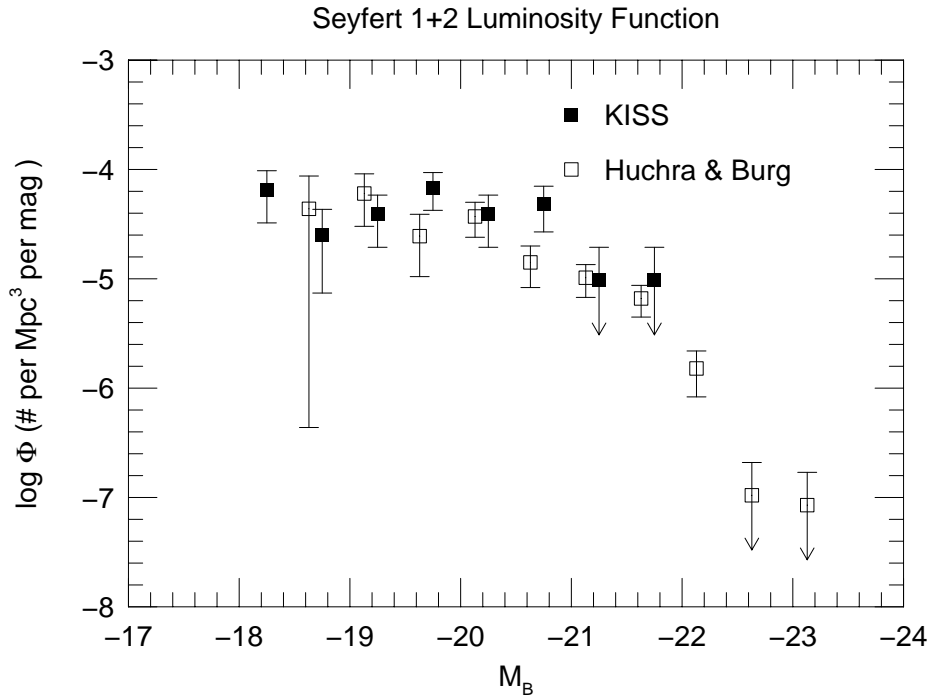}
\caption{Integrated blue luminosity function for Seyfert galaxies
in the KISS AGN sample.  The KISS luminosity function 
is shown in solid squares while
the luminosity function from Huchra \& Burg (1992) is shown with
open squares.}
\end{figure}

\subsection{Multiwavelength Properties}

We have also cross-correlated our H$\alpha$-selected sample
of ELGs from both our first and second survey strips with surveys
done in the x-ray, radio, and the far-infrared.  
A comparison of the KISS sample with the ROSAT All-sky Survey 
(Voges et al. 1999)
finds that 17 of the 2158 KISS ELGs are detected by ROSAT.
These are all AGNs:  13 Sy 1's, 2 Sy 2's, and 2 LINERs.  The
x-ray detected AGNs are primarily the most luminous AGNs detected
by KISS.
A cross-correlation of the
KISS sample with the 1.4 GHz FIRST survey (White et al.~1997) 
done with the VLA reveals that 178 of the 2158 KISS ELGs are
detected by FIRST.  We have follow-up spectra for all of these
sources and find 5 Sy 1's, 28 Sy 2's, and 30 LINERs.  Similarly,
154 of the 2158 KISS ELGs are found in the IRAS Faint Source
Catalog (Moshir et al.~1992) and follow-spectra for these sources
have detected 1 Sy 1, 9 Sy 2's, and 12 LINERs.  The KISS ELGs
follow the well-known radio-IR correlation discovered by 
Condon et al. (1991).
This correlation is thought to be due to the fact that both the
radio and far-IR emission are dominated by star formation.  Somewhat
surprisingly, we find that the majority of the AGNs in our sample
follow the same radio-IR correlation as the star-forming galaxies,
implying that the dominant ionizing source in these objects might be young
stars, not the active nucleus.  We are currently investigating this
possiblity.  We also plan to use these data
to test the validity of IR color diagnostics for AGN activity.

\section{Summary}

The KISS sample represents a substantial step forward in our 
understanding of the AGN population in the local universe.  
It is essentially volume-limited and less biased against redder 
Seyfert 2 and LINER galaxies, and provides a well-defined sample
of nearby ($z < 0.1$) AGN with which to study the statistical
properties of AGN.  In the first KISS survey strip covering
62 square degrees, 10 Seyfert 1's, 19 Seyfert 2's, and 38 LINER's
were discovered.
KISS provides an essential database of
local AGNs for comparison to higher redshift samples.

\acknowledgments
CG is grateful to the American Astronomical Society and the International
Astronomical Union for travel grants which allowed her to participate
in this meeting.  She would also like to thank the Local Organizing
Committee for their wonderful hospitality.

\end{document}